# Domain wall-gated skyrmion logic

Xiangjun Xing,[1,2] Philip W. T. Pong,[1] and Yan Zhou[3,4,*]

[1] *Department of Electrical and Electronic Engineering, The University of Hong Kong, Hong Kong, China*

[2] *College of Physics and Electronic Information Engineering, Wenzhou University, Wenzhou 325035, China*

[3] *Department of Physics, The University of Hong Kong, Hong Kong, China*

[4] *School of Electronic Science and Engineering and Collaborative Innovation Center of Advanced Microstructures, Nanjing University, Nanjing 210093, China*

**ABSTRACT**

Skyrmions and domain walls are typical spin textures of significant technological relevance to magnetic memory and logic applications, where they are used as carriers of information. The unique topology of skyrmions makes them to display distinct dynamical properties compared to domain walls. Some studies have demonstrated that the two topologically inequivalent magnetic objects could be interconverted by cleverly designed geometric structures. Here, we numerically address the skyrmion–domain wall collision in a magnetic racetrack by introducing relative motion between the two objects based on a specially designed junction. An electric current serves as the driving force that moves a skyrmion toward a trapped domain-wall pair. We observe different types of collision dynamics by changing the driving parameters. Most importantly, the domain wall modulation of skyrmion transport is realized in this system, allowing a set of domain wall-gated logical NOT, NAND, and NOR gates to be constructed. By providing a promising logic architecture that is fully compatible with racetrack memories, this work is expected to speed up the development of skyrmion-based magnetic computation.

---

[*] Author to whom correspondence should be addressed: Y. Zhou (E-mail: yanzhou@hku.hk)

## INTRODUCTION

Skyrmions[1–3] and chiral domain walls (DWs)[4,5] are two different kinds of (meta)stable spin configurations in ultrathin magnetic multilayer nanostructures, where the interfacial Dzyaloshinskii-Moriya interaction (DMI)[6–8]—induced conjunctly by the broken inversion symmetry of a ferromagnetic layer and the strong spin-orbit coupling in a heavy-metal layer[9]—tends to twist adjacent magnetic moments and relate a specific chirality to the induced spin configurations. Skyrmions—firstly postulated by Skyrme in a continuous field theory to describe baryon stability[10] and subsequently found in a variety of physical systems[11,12]—carry a conserved topological charge and thus belong to topologically protected structures, that is, they cannot be unwound via continuous deformation[13], which allows them to readily move under an ultralow current density[14–17] and to flexibly avoid pinning centers during motion[18–20]. Apart from these intriguing features, *i.e.*, the flexibility around defects and the high sensitivity to currents, recently, it was established that moving magnetic skyrmions can bring about an emergent electromagnetic field,[15] and most recently, it was demonstrated that magnetic skyrmions can even be controllably written and erased using a tiny mechanical force exerted by a microtip.[21] Because of these merits, magnetic skyrmions serve as an active platform for searching new phenomena[20,22–25] and more importantly as a promising object for building functional devices[18,26–33]. Compared to conventional DWs, chiral DWs show greatly enhanced mobility owing to the combined action of the DMI and the spin-Hall effect, and thus hold great promise for implementing new families of spintronic devices.[34–37]

Up to now, most proposed examples of racetrack memory and logic devices employ skyrmions or DWs as information carriers;[18,26–28,32,33] few devices have been explored which take advantage of potential interplays between skyrmions and DWs for functional operations. In a previous work[29], Zhou *et al.* proposed a hybrid racetrack memory based on coexisting skyrmions and DWs, where skyrmions, directly used for functional operation, are converted from chiral DW pairs through a lateral junction. As revealed in ref.[38], when two skyrmions approach each other, they will shrink in size as a result of mutual repulsion. We now wonder what will happen when a skyrmion collides with a chiral DW. To this end, we need to find, first of all, a system that allows a skyrmion to approach a magnetic DW. Unfortunately, no such system has yet been established so far, although the static structures and dynamic properties of skyrmions and chiral DWs have been well understood[4,13,37,39–41] and the coexistence of these objects in a chiral magnet has been demonstrated[29,42].

It is well established that, in magnetic racetracks, the relation $v_x^d \propto (\beta/\alpha)J$ holds for a skyrmion in steady flow,[19,43,44] where $v_x^d$ is the skyrmion drift velocity along the long axis, $\alpha$ is the damping constant of the used magnetic material, $\beta$ is the ratio of the nonadiabatic and adiabatic spin torques, and $J$ is the current density. Thus, one can obtain higher skyrmion velocity simply by increasing the current density in a given nanotrack. To render a collision, there must be relative motion between a skyrmion and a DW pair, which can in principle be fulfilled by keeping one of them moving while the other at rest. Two pathways are possible: first, one can apply a current small enough to the nanotrack such that the skyrmion moves steadily when the DW is still stationary, and alternatively, one can design a potential profile to confine the DW and leave the skyrmion being free to move. The first scheme has a serious shortcoming, that is, fast skyrmion motion cannot be attained because of the low current density.[18] In turn, the skyrmion might be unable to approach the DW close enough. By contrast, the second scheme does not restrict the current density, and hence high skyrmion velocities are accessible. Comparatively, it is easier to confine a DW using geometric defects.[45] As such, we design an H-shaped junction with four protruding vertices (Fig. 1a) to trap a DW pair (Figs. 1b&2) via pinning effect. We choose an electric current to drive the skyrmion and exploit the resulting collision dynamics of the skyrmion and DW, with several $\beta/\alpha$ values being considered in simulations to account for the divergent results about $\beta$ reported experimentally.[46–49]

We deliver the dynamics of a single skyrmion (Fig. 3) and a single DW pair (Fig. 4) and the collision dynamics of the skyrmion and DW pair (Figs. 5–7) with the current being applied along the nanotrack. It is seen that, in a certain range of $J$, the single skyrmion moves smoothly along the track and passes freely through the junction, whereas the single DW pair, albeit slightly distorted, remains trapped inside the junction, and for the coexisting skyrmion and DW pair system, the skyrmion can get close to the DW pair for collision. By simply varying $J$, we find a series of motional modes for the colliding skyrmion and DW pair. In this system, the skyrmion behaves somewhat like a "rigid ball" and the DW pair like an "elastic cord". Especially intriguing is a dynamic mode raised by lower current densities, under which the skyrmion firstly moves toward and eventually is stopped ahead of the junction by the pinned DW pair. Using this character, we build a skyrmionic logic-NOT gate and develop logical NAND and NOR gates by assembling two NOT gates in parallel and in series, respectively (Fig. 8). Complex circuits, capable of performing arbitrary logic operations, are expected to be implemented by combining these basic components (*i.e.*, NOT, NAND, and NOR). This DW-gated skyrmion logic architecture, together with racetrack memory[28,29], might open new opportunities for skyrmion-based computation.

**MATERIALS AND METHODS**

The nanotrack is a laterally confined ultrathin multilayer film (Fig. 1a) with asymmetric interfaces to engender an interfacial DMI[8]. A skyrmion and a DW pair (Fig. 1b) are written into the nanotrack using perpendicular currents across two separate nanocontact spin valves at the positions denoted by the elements in yellow (Fig. 1a). The H-shaped junction is used mainly to stabilize the DW pair immersed in an in-plane current, although it can also assist the nucleation of the DW pair initiated by a perpendicular current. In the following, we will describe, firstly, the injection process of a DW pair under a vertical current (Fig. 2). The injection of a skyrmion with a perpendicular current is not covered here, because it has already been numerically delivered in refs.[19,30,50]. Subsequently, we will examine successively the motion dynamics of a skyrmion, a DW pair, and the coexisting skyrmion and DW pair triggered by an in-plane current. For these studies, the nanotrack used is 600 nm long and 60 nm wide (see Fig. 1b for the definition of geometric parameters). Finally, we will demonstrate the logic NAND and NOR operations. To this end, we assemble single nanotracks to form composite structures. For all simulations, the nanotracks are 1 nm in thickness.

Micromagnetic simulations, using the public-domain GPU code—MuMax3[51], were carried out to unravel the dynamics of the skyrmion and/or the DW pair under electric currents by numerically solving the Landau-Lifshitz-Gilbert equation[52,53] augmented with a spin-torque term, which is either in the Slonczewski form[54] or the Zhang-Li form[55] depending on the current direction (perpendicular or in-plane). For all computations, the interfacial DMI term was incorporated into the conventional LLG equation.[39] Material parameters used in simulations are typical of the Co/Pt multilayer system with perpendicular magnetic anisotropy:[3,19,56] the saturation magnetization $M_s$=580 kAm$^{-1}$, the exchange stiffness $A$=15 pJm$^{-1}$, the damping constant $\alpha$=0.3, and the electron spin polarization $P$=0.4. For the in-plane current, three representative cases of $\beta$=0.5$\alpha$, $\alpha$, and 2$\alpha$ are considered, where $\beta$ stands for the nonadiabaticity efficiency of the Zhang-Li torque. A series of $K_u$ (perpendicular magnetocrystalline anisotropy) and $D$ (DMI strength) combinations were checked in computations to ensure that the obtained results could be applied to a variety of samples.[5,40,42,57] The presented results correspond to $K_u$=0.8 MJm$^{-3}$ (the effective uniaxial anisotropy $K_{eff}$=0.6 MJm$^{-3}$ according to $K_{eff}=K_u-(1/2)\mu_0 M_s^2$) and $D$=3.5 mJm$^{-2}$. The simulation volume, irrespective of its size, is discretized into an array of 1×1×1 nm$^3$ regular meshes for finite-difference computation.

**RESULTS**

*Injection of a confined domain-wall pair*

Figure 2a shows the vertical structure of the nanocontact used to polarize the perpendicular current and shape its path. The current flows across a local area (as enclosed by the yellow box in Fig. 2b) in the Co layer beneath the spacer layer, where the magnetic moments sense a spin torque. The injection process of the DW pair is illustrated in Fig. 2c. Under current action, the magnetic moments at the edges reverse at first forming two edge-merons[58] (Fig. $2c_1$), as revealed by the increasing topological charge (Fig. 2d). Next, the edge-merons expand in size (Fig. $2c_2$) and approach each other generating an antivortex[59] in between (Fig. $2c_3$), when the total energy of the system reaches a peak (Fig. 2e). Further, the two distorted edge-merons touch each other (Fig. $2c_4$)—with a quasi-Bloch-point[60] nucleated in between and rapidly annihilated—before merging into a DW pair (Fig. $2c_5$). In the 20 ps time interval from 170 ps to 190 ps, the exchange energy, total energy, and topological charge are dissipated through releasing the Bloch-point-like object (Fig. 2e). Finally, a stable DW pair is formed within the junction, as indicated by the unchanging magnetization, topological charge, and energies in Fig. 2(d,e).

The applied current must be sufficiently large to overcome the energy barrier associated with the merging of the two edge-merons, and it must be reasonably small to prevent the final DW pair from extending outside the junction by suppressing rapid motion of the magnetic moments directly under the current as well as their strong correlation with other magnetic moments. The edge magnetic moments respond firstly because they have in-plane components (due to the DMI) and thus the spin transfer between these magnetic moments and the polarized electrons is most efficient.

*Current-driven motion of a single skyrmion*

The spin valve for skyrmion injection is analogous in structure as for DW injection except for the difference in lateral geometry; the detailed skyrmion-injection process has been presented in literatures[19,29,30] and thus is not replicated here. In simulations, we preset a bubble-like spin configuration[61] and relax it to the equilibrium skyrmion for dynamics study. Fig. 3a plots the skyrmion motion modes as a phase diagram in the two-dimensional space formed by the current density $J$ and the nonadiabaticaity coefficient $β$. Depending on the values of $J$ and $β$, the skyrmion can pass through the junction (denoted as "Pass" in Fig. 3a), be blocked by the junction (denoted as "Fail" in Fig. 3a), or be ejected from the boundary (denoted as "Fail" in Fig. 3a), as shown sequentially in Fig. 3(c–e). Note here that "Pass" means that the magnetic object, after passing the junction area, can reach the right terminal of the nanotrack; otherwise, the relevant mode is categorized into "Fail".

The reasons why the skyrmion is stopped at the junction for $(J, β)=(1.0×10^{12}$ Am$^{-2}$, 0.15) are: firstly, the junction's inclusion modifies the energy landscape[45] of the nanotrack and introduces local potential wells around the junction vertices, and, secondly, the skyrmion moves toward the side boundary due to skyrmion Hall effect[24,62] (*i.e.*, a skyrmion moves transversely because of the Magnus force [$F_g ∝ (1−β/α)J$] associated with its longitudinal motion[19,43]) and drops into the potential well and cannot escape from it because of the small spin torques (adiabatic term $τ_a ∝ J$ and nonadiabatic term $τ_b ∝ βJ$;[55] therefore both are the smallest among the cases considered in Fig. 3a). By contrast, skyrmion expulsion at $(J, β)=(4.0×10^{12}$ Am$^{-2}$, 0.6) is simply caused by the skyrmion Hall effect and is independent of the junction's presence. In other words, assuming the same ($J$, $\beta$) values, a skyrmion would still be annihilated at the boundary even for a smooth nanotrack.

The skyrmion position versus time curves (Fig. 3f) can provide more information about the skyrmion motion dependent on ($J$, $\beta$). The skyrmion under ($J$, $\beta$)=(1.0×10$^{12}$ Am$^{-2}$, 0.15) moves at a constant speed initially and stops eventually when it reaches the junction region as revealed in the snapshot image (Fig. 3d). Once falling into the potential well, the skyrmion is tightly bound there unless a stronger current is applied. The difference between the skyrmion motions under ($J$, $\beta$)=(2.0×10$^{12}$ Am$^{-2}$, 0.15) and (1.0×10$^{12}$ Am$^{-2}$, 0.15) is that for the former ($J$, $\beta$) the skyrmion can escape from the potential well, because the current's driving force exceeds the well's restoring force. For ($J$, $\beta$)=(2.0×10$^{12}$ Am$^{-2}$, 0.3), no Magnus force acts on the skyrmion [because $F_g \propto (1-\beta/\alpha)J=0$ for $\beta=\alpha$], so that the skyrmion moves along the middle axis of the nanotrack when it is far from the right end (Fig. 3c), resulting in a constant slope in the position versus time curve (Fig. 3f). The situation for ($J$, $\beta$)=(3.0×10$^{12}$ Am$^{-2}$, 0.3) is the same as for (2.0×10$^{12}$ Am$^{-2}$, 0.3) except for the increased skyrmion velocity. The skyrmion ejection at ($J$, $\beta$)=(4.0×10$^{12}$ Am$^{-2}$, 0.6) manifests itself as vanishing data points in the position versus time curve after 1.5 ns. We also note that the skyrmion slows down after it enters the junction region and recovers to its initial speed once it leaves the junction. The skyrmion deceleration inside the junction originates from the dragging of the potential well. When $J$ is decreased to 2.0×10$^{12}$ Am$^{-2}$ ($\beta$ kept at 0.6), the skyrmion only sees negligible influence of the junction, as revealed by the uniform slope of the position versus time curve. In this case, the skyrmion trajectory is slightly away from the middle axis of the nanotrack since the Magnus force is small [recall that $F_g \propto (1-\beta/\alpha)J$;[19,43] $J$ is small despite $1-\beta/\alpha \neq 0$].

Based on Fig. 3, we can conclude that the skyrmion motion behaviors in the junction-contained nanotrack are much similar as in a straight nanotrack for most cases excluding $(J, \beta)=(1.0\times10^{12}\text{ Am}^{-2}, 0.15)$.

*Current-driven dynamics of a confined domain-wall pair*

Figure 4 represents the results for current-driven dynamics of a confined DW pair. As seen from Fig. 4a, for most $(J, \beta)$ combinations, the DWs fail to escape from the confinement potential of the junction. Among the considered cases, only for the largest $(J, \beta)$ [*i.e.*, $(4.0\times10^{12}\text{ Am}^{-2}, 0.6)$] can the DW pair succeed in passing through the junction area. Fig. 4c shows the DW pair under $(J, \beta)=(2.0\times10^{12}\text{ Am}^{-2}, 0.3)$. It indicates that the DW pair structure is slightly modified immediately after current application (*e.g.*, 0.2 ns; Fig. 4c$_1$), and no more variation occurs to the structure from 0.2 ns to 10 ns (compare Fig. 4c$_1$ and Fig. 4c$_2$), implying that the DW pair cannot be released by small driving forces, even for long action time. For $(J, \beta)=(4.0\times10^{12}\text{ Am}^{-2}, 0.6)$, the DW pair is firstly bent and then twisted (Fig. 4d$_1$), resulting in an asymmetric DW profile with respect to the nanotrack's middle axis, which can be attributed to the DMI.[29,40,41] Subsequently, the near and far DWs coalesce at around the upper boundary, where they meet because of the heavy distortion. In this way, a DW pair, with two open ends pinned at two separate boundaries, is depinned from one boundary, resulting in an intermediate magnetic object with one open end, as shown in Fig. 4d$_2$.

After escaping from the junction, the magnetic object converts into an edge-meron[58], which flows steadily along the nanotrack boundary and ultimately leaves the nanotrack at the right terminal. The complex evolution process from the DW pair to the edge-meron manifests itself as the changing topological charge and energies of the system (Fig. 4e).

By comparing Figs. 3a&4a, one can find that for every ($J$, $β$) combination—under which the single skyrmion can pass the junction, the DWs can be tightly trapped inside the junction.

*Current-driven collision dynamics of skyrmion and domain-wall pair*

Now, we turn to addressing the collision dynamics of the coexisting skyrmion and DW pair in response to an in-plane current. Whether the trapped DWs can stop any magnetic objects (the original skyrmion or other secondary objects) passing through the junction relies on the current parameters, *i.e.*, ($J$, $β$), as shown in Fig. 5a. One can see that for lower values of ($J$, $β$), no magnetic objects can penetrate the junction. Only when $J$ and $β$ both become sufficiently large, can some magnetic objects enter into the right branch and move steadily along the nanotrack's right branch. Next, we detailedly analyze the collision processes of the skyrmion and DWs.

For ($J$, $β$)=($1.0×10^{12}$ Am$^{-2}$, 0.3), the skyrmion is finally blocked in front of the DW pair, as shown in Fig. 5c. Here, the current-induced force cannot overcome the repulsion force from the near DW as the skyrmion moves close to it, and eventually the two forces come into balance at a certain distance between the skyrmion and the near DW. Increasing $J$ while keeping $β$ unchanged will lead the skyrmion to be expelled from the boundary (Fig. 5d). The detailed process is as follows: as the skyrmion gets much closer to the near DW, the repulsion force between them further distorts the DW pair, so that the asymmetry[40,41] of the DW pair relative to the long axis of the nanotrack is enhanced and in turn leads to a sizable transverse (*i.e.*, *y*-axis directed) force on the skyrmion. As a result, the skyrmion moves transversely (Fig. 5d$_1$), touches the lowerleft vertex of the junction (Fig. 5d$_2$), and finally be ejected. Under the repulsion force of the far DW, the near DW restores and even extends beyond the junction (Fig. 5d$_3$). However, it is still strongly pinned.

The dynamic processes in Fig. 5(c,d) are apparently reflected on $Q(t)$ and $E(t)$ curves in Fig. 5(e,f). From Fig. 5e, we notice that $Q$ changes only slightly with time, which is because no drastic structural variation happens to the skyrmion or DW pair during current action, the net effect of which is the reduced spacing between the skyrmion and the near DW. The jump in exchange energy before 4 ns corresponds to the increasing repulsion between the skyrmion and near DW. Nevertheless, from Fig. 5f, we see a sharp decrease in $Q$ and $E$ at ~2.73 ns (Fig. 5d$_2$), which characterizes the breakdown of the skyrmion structure at the boundary.[43]

Further increasing the current density while maintaining *β* at 0.3 can cause the DW pair to partially detach from the junction. The whole process is depicted in Fig. 6a, where the initial process before 1.6 ns is similar to that before 2.7 ns for (*J*, *β*)=(2.0×10$^{12}$ Am$^{-2}$, 0.3); compare Fig. 6c with Fig. 5f. At this stage, the skyrmion approaches the DW pair enabling mutual repulsion; consequently, the DW pair distorts and the skyrmion shrinks, deviates from the nanotrack's middle axis, and ultimately touches the boundary becoming a fractional skyrmion[29]. After 1.6 ns, the fractional skyrmion continues to shrink in size, and remarkably, the DW pair, enclosed by two open strings, is depinned from the lower boundary, becoming the other fractional skyrmion encircled by a single string (Fig. 6a$_5$), for which $Q$ is ~0.65 (Fig. 6c). Under current action, this fractional skyrmion reaches a dynamic equilibrium configuration, which is maintained by three independent forces, *i.e.*, $\mathbf{F}_v$, $\mathbf{F}_g$, and $\mathbf{F}_p$, as shown in Fig. 6a$_6$. Once the current is switched off, the fractional skyrmion decays rapidly and is ejected after ~0.6 ns from the boundary (see Fig. S1 for detail), since the force balance between $\mathbf{F}_v$, $\mathbf{F}_g$ and $\mathbf{F}_p$ is broken. On the other hand, once the current density increases to $4.0\times10^{12}$ Am$^{-2}$, the fractional skyrmion is pulled into the nanotrack interior becoming a skyrmion (see Fig. S2).

*Force analysis.* The forces—experienced by the fractional skyrmion in equilibrium—should satisfy the Thiele equation that reads,[17,19,43,44,63]

$$\mathbf{G}\times(\mathbf{v}^s-\mathbf{v}^d)-\nabla V+\mathcal{D}(\beta\mathbf{v}^s-\alpha\mathbf{v}^d)=0,$$

which describes the balance of the Magnus force [$\mathbf{F}_g=\mathbf{G}\times(\mathbf{v}^s-\mathbf{v}^d)$], the confining force ($\mathbf{F}_p=-\nabla V$), and the viscous force [$\mathbf{F}_v=\mathcal{D}(\beta\mathbf{v}^s-\alpha\mathbf{v}^d)$]. $\mathbf{G}=G\hat{e}_z$ is a gyrocoupling vector with $G$ being proportional to $Q$ and $\hat{e}_z$ representing the unit vector along $z$ axis, $V$ is the confining potential due to boundaries, and $\mathcal{D}=\begin{pmatrix}\mathcal{D}_{xx} & \mathcal{D}_{xy}\\ \mathcal{D}_{yx} & \mathcal{D}_{yy}\end{pmatrix}=\begin{pmatrix}\mathcal{D} & 0\\ 0 & \mathcal{D}\end{pmatrix}$ is a dissipation tensor. $\mathbf{v}^d$ is the drift velocity of a spin texture, and $\mathbf{v}^s=-[\gamma\hbar P/(2\mu_0 eM_s)]J\hat{e}_x$ is the velocity of conduction electrons,[55] where $\hat{e}_x$ is the unit vector along $x$, $\gamma$ the gyromagnetic ratio, $\mu_0$ the vacuum permeability, $\hbar$ the reduced Planck constant, and $e$ the elementary charge. Here, the fractional skyrmion is static (dynamically stabilized), *i.e.*, $\mathbf{v}^d=0$; therefore, $\mathbf{F}_g = \mathbf{G} \times \mathbf{v}^s$ and $\mathbf{F}_v = \mathcal{D}\beta\mathbf{v}^s$, which allow us to identify the directions of the two forces (as labeled in Fig. 6a$_6$). In the present case, the confining force $\mathbf{F}_p$ arises sorely from the boundaries and acts as a passive force to counteract the Magnus and viscous forces, leading to zero net force upon the fractional skyrmion.

When the current density increases further such that $(J, \beta)=(4.0\times10^{12}$ Am$^{-2}$, 0.3), the DW pair converts into a skyrmion, as illustrated in Fig. 6b. The process of the first 1 ns still resembles the initial processes described above [compare Fig. 6(a$_2$,b$_2$)]. Because of the increased current density, the skyrmion gets closer to the DW pair, making the latter to undergo stronger distortion. As a result, the two DWs merge at the upper boundary forming a fractional skyrmion (#2 in Fig. 6b$_3$), different from what occurs in Fig. 6a, and meanwhile, the skyrmion touches the lower boundary giving the other fractional skyrmion (#1). Next, as the fractional skyrmion #1 continues to move rightward, the fractional skyrmion #2 is forced to depin from the junction and transform into an elongated magnetic bubble[19,29] (Fig. 6b$_4$). Subsequently, the fractional skyrmion #1 leaves the nanotrack and the bubble (#2) relaxes into a regular skyrmion. Thereafter, the new skyrmion moves along the nanotrack's middle axis and melts at the right terminal. The two events, *i.e.*, the conversion from a DW pair to a skyrmion and the annihilation of the new skyrmion, manifest themselves as two abrupt drops in the $Q(t)$ curve of Fig. 6d, from ~1.5 to 1.0 and 1.0 to 0, respectively.

In Figs. 5 and 6, the dynamic processes are presented detailedly for $\beta=0.3$. Now, we go into the details for $\beta=0.6$ (Fig. 7). For $(J, \beta)=(3.0\times10^{12}$ Am$^{-2}$, 0.6), because of a Magnus force, the skyrmion deviates from the nanotrack's middle axis[19,43] towards the upper boundary in approaching the DW pair (Fig. 7a$_1$). The DW pair depins from the junction at the upper boundary (Fig. 7a$_2$) in response to the repulsion force from the skyrmion. The newly generated fractional skyrmion no longer detaches from the lower boundary as in Fig. 6(a,b), where the repulsion force from the annihilating skyrmion dominates the DW pair depinning. In this case, the complete skyrmion still contributes a repulsion force, which tends to expel the new fractional skyrmion from the nanotrack. Since $\beta \neq \alpha$, a Magnus force happens to the fractional skyrmion and counterbalances the repulsion force from the skyrmion, so that the fractional skyrmion is dynamically protected from being ejected and converts into an edge-meron[58] (Fig. 7a$_4$). After passing the junction, the skyrmion and the edge-meron moves steadily in the right branch and reach the right terminal [Fig. 7(a$_5$–a$_8$)]. The drops in $Q(t)$ curve in Fig. 7c around 1.3 ns, 2.6 ns, and 3.2 ns denote the edge-meron depinning from the junction, the edge-meron annihilation, and the skyrmion annihilation at the right terminal, respectively.

For $(J, \beta)=(4.0\times10^{12}$ Am$^{-2}$, 0.6), there is no interaction between the skyrmion and the DW pair because of a large distance between them. Thus, the motions of the two objects are fully independent. The motions of the coexisting skyrmion and DW pair (Fig. 7b) appear to be a superposition of the individual motions of the single skyrmion and the single DW pair (Figs. 3&4).

The behaviors of the coexisting skyrmion and DW pair can be briefly summarized as follows. For $(J, \beta)=(1.0\times10^{12}$ Am$^{-2}$, 0.3), the skyrmion is blocked in front of the junction by the DW pair, although the latter is slightly modified in shape.

For $(J, \beta)=(2.0\times10^{12}$ Am$^{-2}$, 0.3), the skyrmion is stopped ahead of the junction and ejected from the lower boundary, and the DW pair is distorted. For $(J, \beta)=(3.0\times10^{12}$ Am$^{-2}$, 0.3), the DW pair blocks the skyrmion and pushes it out of the nanotrack, and eventually the DW pair is converted into a fractional skyrmion stabilized dynamically. For $(J, \beta)=(4.0\times10^{12}$ Am$^{-2}$, 0.3), the DW pair still blocks the skyrmion by pushing it out of the nanotrack, but meanwhile the DW pair is entirely depinned from the junction and becomes a skyrmion. For $(J, \beta)=(3.0\times10^{12}$ Am$^{-2}$, 0.6), the DW pair no longer blocks the skyrmion and is forced to depin from the junction becoming an edge-meron. For $(J, \beta)=(4.0\times10^{12}$ Am$^{-2}$, 0.6), both the skyrmion and the DW pair overcome the barrier of the junction and enter the right arm, although the skyrmion annihilates immediately at the upper boundary and the DW pair converts into an edge-meron.

The rich collision dynamics of the coexisting skyrmion and DW pair embodies the complex competition among the current-induced forces[19,43] (*i.e.*, viscous force and Magnus force), the confining force due to boundaries, and the repulsive force between magnetic objects. Approximately, the higher the current density is, the nearer the skyrmion can approach the DW pair and in turn the stronger the repulsion between the skyrmion and the DW pair. The ejection of the skyrmion from the nanotrack shown in Figs. 5d&6 is primarily due to the repulsion force other than the Magnus force, which however is in charge of the skyrmion ejection in Fig. 3e. The strength and direction of the repulsive force are determined by the shapes, sizes, positions, and spacing of the skyrmion and the DW pair, which are rapidly evolving under the simultaneous action of the aforementioned forces. Furthermore, as presented above, the Magnus and viscous forces are a function of the drift velocity ($\mathbf{v}^d$) of a magnetic object, and the confining force relies on the shape, size, and position of a magnetic object.[19,43,63] Therefore, *the forces and the motional states are mutually dependent*, which makes it impossible to analytically address the collision dynamics of the coexisting skyrmion and DW pair.

### *Logic operations by domain-wall-mediated skyrmion motion*

Now, let us neglect the detailed motion processes of the skyrmion and the DW pair and simply concentrate on the blocking effect of the DW pair on the skyrmion by comparing Figs. 3a,4a&5a. Clearly, for most combinations of ($J$, $\beta$), if no DW pair is contained in the junction, a skyrmion can pass through the junction; if a DW pair is present in the junction, a skyrmion cannot pass through the junction, since it is either stopped in front of the DW pair or pushed out. This is indeed equivalent to a logic-NOT operation,[27] for which the presence of the DW pair (skyrmion) in the junction (right branch) of the nanotrack is encoded into "1" at the input (output) terminal and their absence into "0". In this logic-NOT gate, the skyrmion behaves as a carrier of information and the DW pair as a gatekeeper that controls the information flow.

A logic-NOT gate is a 1-input signal-processing component; nevertheless, before any arbitrary logical functions can be implemented, at least one 2-input gate has to be constructed preliminarily.[27] By virtue of the structural characters of the logic-NOT gate, we propose two types of such components—NAND and NOR gates[64–66]. The schematic architectures are shown in Fig. 8. The NAND gate is built from two NOT gates connected in parallel, and the NOR gate is based on two serially connected NOT gates. For each 2-input gate, the two inputs are encoded into the individual magnetic states of the two junctions, and the output is encoded into the magnetic state of the right arm.

We numerically test the functionality of the proposed 2-input gates with ($J$, $β$)=(2.0×10$^{12}$ Am$^{-2}$, 0.3), and the operation results for the NAND and NOR gates are illustrated in Fig. 8(a–c) and Fig. 8(d–f), respectively. The left panels show the initial states for each operation. A current is switched on at 0 ns to move the skyrmion carrier, and for the NAND (NOR) gate the magnetic state at the output port is measured at 3.5 ns (5.5 ns). Once one or more skyrmions are detected, the output will be recognized as "1"; otherwise, as "0". The detection can be achieved using GMR effect or by virtue of the emergent electrodynamics intrinsic to skyrmions. Using this coding scheme, the truth tables for the two gates are derived. Note that the operation details of 1|0→1 for NAND and 1|0→0 for NOR are omitted here, since they resemble those of 0|1→1 and 0|1→0, respectively.

**DISCUSSION**

Scrutinizing the collision processes, one can find that even for $β$=0.3, the skyrmion always deviates from the middle axis once it is close to the near DW. However, in this case no transverse force can emerge from a current since $β=α$ leads to zero Magnus force.[19,43] Therefore, the transverse skyrmion motion should be due to a transverse component of the repulsion force, which can be ascribed to the asymmetry in the DW profile.[40,41]

Geometrically, a closed string must not be in contact with any a boundary, and once a string intersects a boundary, it is split apart at the contact point. In other words, if two DW-ends merge, the merging point must lie at the boundary. Therefore, from the string-geometry point of view[29], the DW pair after collision can only exist as one of the three configurations, namely, a slightly modified DW pair, a fractional skyrmion (edge-meron), or a skyrmion. In simulations, we indeed observe all the three states, which are dynamically enabled by applying various collision conditions set by the current parameters—$J$ as well as $\beta$. Regarding the blocking effect of the DW pair, there also exist three outcomes, namely, the skyrmion is stopped and the DW pair remains trapped inside the junction, which situation prevails for most ($J$, $\beta$) combinations considered, or the skyrmion is obstructed and the DW pair is released, or the skyrmion passes the junction and the DW pair detaches.

There is also repulsion between the two DWs forming the pair, especially when the near DW is driven close to the far DW and/or when it is distorted by a force either from a current or from the skyrmion. During collision, the skyrmion is always rigid and well preserves its shape (*i.e.*, without deformation) with its size being gradually reduced in approaching the DW. It can only be corrupted through intersecting geometric boundaries.[38,41] On contrary, the DWs are rather 'soft'; they deform in response to the forces the current and/or the skyrmion exert. Owing to the softness of the DW pair, there is no restriction on the geometric configuration of the evolving DW pair, so that multiple motion modes can be dynamically activated for the coexisting skyrmion and DW pair by varying current densities. Here, the topological property of a spin texture manifests itself again. The skyrmion is protected by its quantized topological charge from being distorted[13,20] regardless the current parameters, whereas the DW pair, lacking such protection, is twisted, stretched, and/or merged depending strongly on the current parameters.

    We would like to mention that, Kunz[67] numerically addressed *magnetic field-driven collision of transverse DWs* in soft magnetic nanowires, which was found to be able to annihilate DWs or render a 360° DW depending on the relative alignments of the DWs, and furthermore he demonstrated collision-assisted DW depinning from a notch under an eightfold smaller field. In a seminal work of 2005, Allwood *et al.*[27] experimentally demonstrated a set of DW-based logic gates operated using pulsed magnetic fields, which has inspired enormous research activities on both fundamental and technological issues of DWs. In the long term, DWs might have to eventually give way to skyrmions for magnetic memory and logic applications, since the latter promises denser and more energy-efficient devices.[18] Recently, our group[32] demonstrated another two kinds of skyrmion-based, 2-input logical gates, that is, AND and OR, which are based on a skyrmion-merging architecture and would require sophisticated procedures for practical fabrication. In a later work[33], we proposed a voltage-gated skyrmion transistor, which possesses a simple structure and can indeed behave as a logical NOT gate, if the input and output are encoded into the control voltage and local magnetization orientation, respectively. Despite the similarity in circuit structure to the voltage-gated gate, the NOT and resulting NAND and NOR

gates demonstrated here operate under distinct gating mechanism that originates from the skyrmion–DW interplay.

Magnonics[68] (*i.e.*, magnon spintronics), as a rapidly growing research field, represents another technological route for circumventing the bottleneck of today's CMOS-based electronics. The prominent advantages that magnonic circuitry can offer include the substantially enhanced throughput and energy efficiency resulting from the inherent high group velocity and low power of spin waves.[69,70] In this aspect, it is difficult for skyrmion-based devices to outperform spin wave-based ones unless ultrahigh skyrmion mobility can be acquired, for example, by improving spin transfer efficiency between driving currents and local magnetic moments. Compared to spin-wave logic[64–66], skyrmion logic exhibits better compatibility with racetrack memory[18,26,28] and thus makes hardware reconfiguration easier, since the former uses magnons rather than skyrmions as information carriers.

It is worth noting that in this proof-of-principle demonstration of DW-gated skyrmion logic, the Zhang-Li form of spin torques is used to trigger skyrmion movement. However, in practical implementation, the device structure and materials can be optimized to benefit from the high efficiency of emerging spin-orbit torques,[71–73] promising energy-efficient operations at far reduced current densities. This study helps uncover the difference in the dynamic properties of topological and topologically trivial magnetic objects, which are inaccessible by simply using external stimuli such as electric currents.

**CONCLUSION**

In conclusion, we realize current-driven skyrmion–DW collision by introducing a planar junction into a magnetic racetrack, which can effectively trap a DW pair while allowing a skyrmion to freely pass through. Using this structure, we systematically address the collision dynamics of coexisting skyrmion and DW pair, and especially, we identify DW-modulated skyrmion motion. Finally, we propose the concept of DW-gated skyrmion logic by demonstrating a set of elementary logic gates.

**Acknowledgements:** Y.Z. acknowledges the support by National Natural Science Foundation of China (Project No. 1157040329), the Seed Funding Program for Basic Research and Seed Funding Program for Applied Research from the HKU, ITF Tier 3 funding (ITS/203/14), the RGC-GRF under Grant HKU 17210014, and University Grants Committee of Hong Kong (Contract No. AoE/P-04/08). X.J.X. thanks the support by the Zhejiang Provincial Natural Science Foundation of China under Grant No. LY14A040006 and the National Natural Science Foundation of China under Grant No. 11104206.

**Author Contributions:** X.J.X conceived the idea and performed micromagnetic simulations. Y.Z. coordinated and supervised the project. All authors discussed the results. X.J.X. wrote the manuscript.

**FIGURE CAPTIONS**

Figure 1. **Sketch of device structure.** (**a**) Perspective view of the magnetic racetrack with an H-shaped junction. Skyrmions and DWs can be injected into the nanotrack using nanocontact spin valves situated at the yellow boxes. (**b**) A DW pair can be trapped in the junction, when a moderate current is applied to the nanotrack, whereas a skyrmion can freely flow along the nanotrack and gradually approach the DW, resulting in a skyrmion–DW collision. The skyrmion–DW repulsion together with current-induced torque can trigger complex collision dynamics. The nanotrack is 60 nm wide and 1 nm thick. The junction size *a*, *b*, and *c* are 20 nm, 10 nm, and 30 nm, respectively.

Figure 2. **Writing a DW pair into the junction**. (**a**) Structure of nanocontact spin valve. (**b**) Initial state of the nanotrack. The yellow box encircles the current action area. (**c**) Injection process for the DW pair. Current action time is specified in each panel. (**d**) Topological charge $Q$ and vertical magnetization $m_z$, averaged over the entire volume of the nanotrack, against current action time. (**e**) Exchange energy $E_{ex}$ and total energy $E_{tot}$ against current action time. The vertical dashed lines in panels **d**, **e** mark the temporal moments specified in panel **c**.

Figure 3. **Current-driven skyrmion motion.** (**a**) Skyrmion behavior versus current parameters. 'Pass' ('Fail') means that the skyrmion can (cannot) pass through the junction region and reach the right terminal of the nanotrack; 'Fail' denotes that the skyrmion cannot pass the junction. (**b**) Initial magnetization state at 0 ns when the current is switched on. (**c**–**e**) Skyrmion configuration at special time after current

turnon. (**f**) Skyrmion position against current action time for individual ($J$, $\beta$) combinations. Zero nanometer and 600 nm correspond to the left and right terminals of the nanotrack, respectively.

Figure 4. **Current-driven DW-pair dynamics.** (**a**) DW-pair behavior versus current parameters. 'Pass' ('Fail') means that the DW-pair can (cannot) depin from the junction and reach the right terminal. (**b**) Initial magnetization state at 0 ns when the current is switched on. (**c, d**) Spin configuration at special time after current turnon. In panel **d**, the DW-pair actually converts into an edge-meron. (**e**) Energies $E_{ex}$, $E_{tot}$ and topological charge $Q$ against current action time. The solid lines are for panel **d**. The dashed line for panel **c**, displaying $Q$ against time, is flat, agreeing with the tiny structural change in the DW pair for ($J$, $\beta$)=($2.0\times10^{12}$ Am$^{-2}$, 0.3).

Figure 5. **Current-driven skyrmion–DW collision dynamics.** (**a**) Blocking effect versus current parameters. 'Pass' means that one or more magnetic objects resulting from the skyrmion–DW collision can pass the junction and reach the right terminal; 'Fail' denotes that no magnetic object can pass through the junction and reach the right terminal. (**b**) Initial magnetization state at 0 ns when the current is switched on. (**c, d**) Spin configuration at special time after current turnon. In panel **c**, the skyrmion is stopped ahead of the DW pair; in panel **d**, the skyrmion is expelled from the lower boundary. (**e, f**) Energies $E_{ex}$, $E_{tot}$ and topological charge $Q$ against current action time. Panels **e**, **f** correspond to panels **c**, **d**, respectively.

Figure 6. **Skyrmion–DW collision at increased current densities with $\beta$=0.3.** (**a**, **b**) Spin configuration at special time after current turnon. In both panels **a**, **b**, the original

skyrmion is ejected from the lower boundary. In panel **a**, the DW pair is partially depinned from the junction, becoming a dynamically stabilized fractional skyrmion; in panel **b**, the DW pair is converted into a new skyrmion. (**c, d**) Energies $E_{ex}$, $E_{tot}$ and topological charge $Q$ against current action time. Panels **c**, **d** correspond to panels **a**, **b**, respectively.

Figure 7. **Skyrmion–DW collision with $\beta$=0.6.** (**a**, **b**) Spin configuration at special time after current turnon. In both panels **a**, **b**, the DW pair is converted into an edge-meron. In panel **a**, the skyrmion passes through the junction and reaches the right terminal; in panel **b**, the DW pair goes through the junction area but melts immediately at the upper boundary. Note that there is no skyrmion–DW interplay in panel **b**; each individual objects respond independently to the current-induced torque. (**c, d**) Energies $E_{ex}$, $E_{tot}$ and topological charge $Q$ against current action time. Panels **c**, **d** correspond to panels **a**, **b**, respectively.

Figure 8. **Layout and operation of logic NAND and NOR functions.** Two single NOT gates connected in parallel or in serial make a NAND or a NOR gate. The purple and orange boxes denote the input and output ports, respectively. (**a**–**c**) Logic NAND gate. (**d**–**f**) Logic NOR gate. The left panels show the initial states before operation, and the right panels give the results after operation. The truth tables are attached on the right. For each operation, a current with $(J, \beta)$=(2.0×10$^{12}$ Am$^{-2}$, 0.3) is used as a driving force. The current action time is 3.5 ns for NAND and 5.5 ns for NOR.

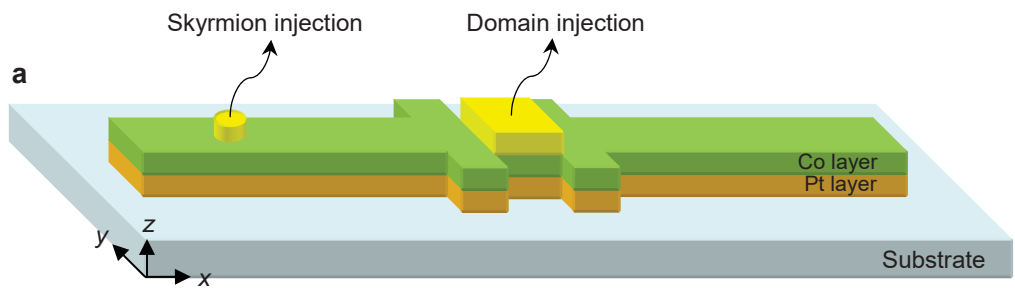
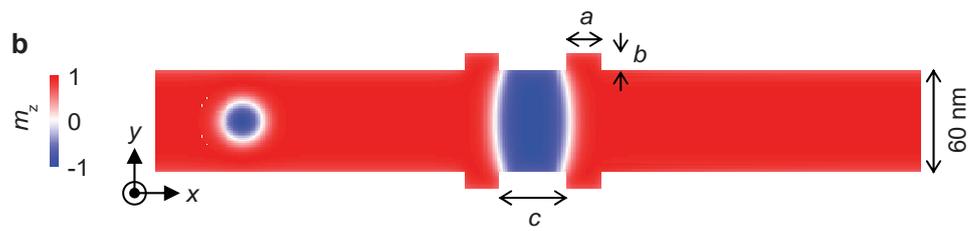

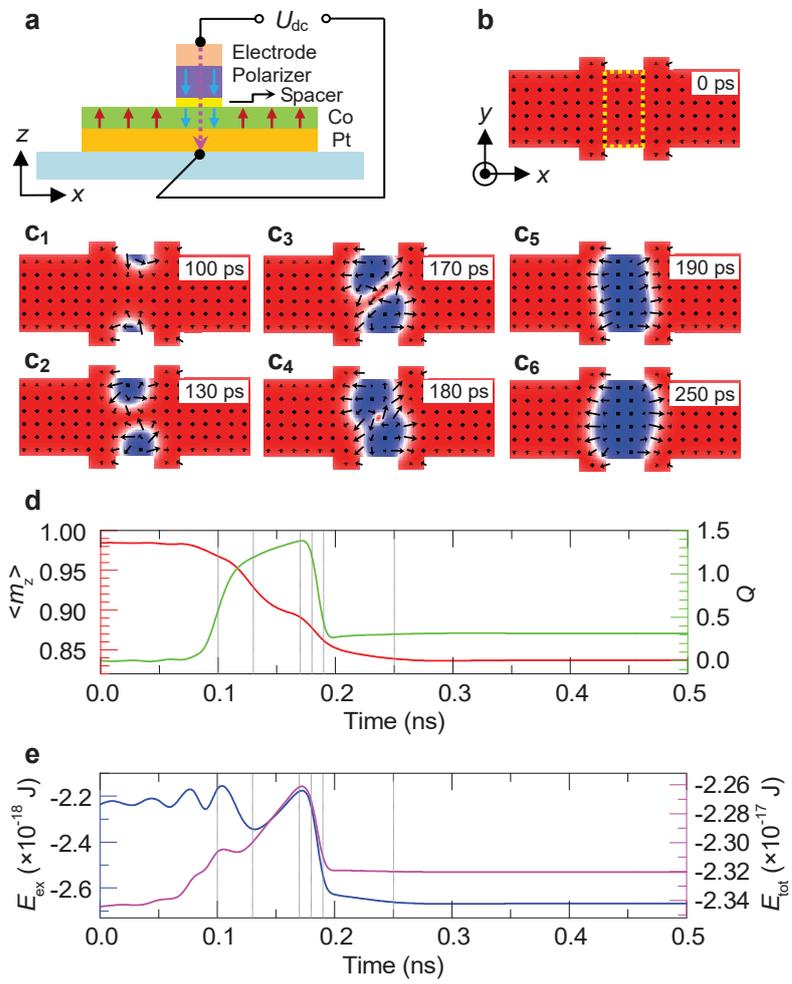

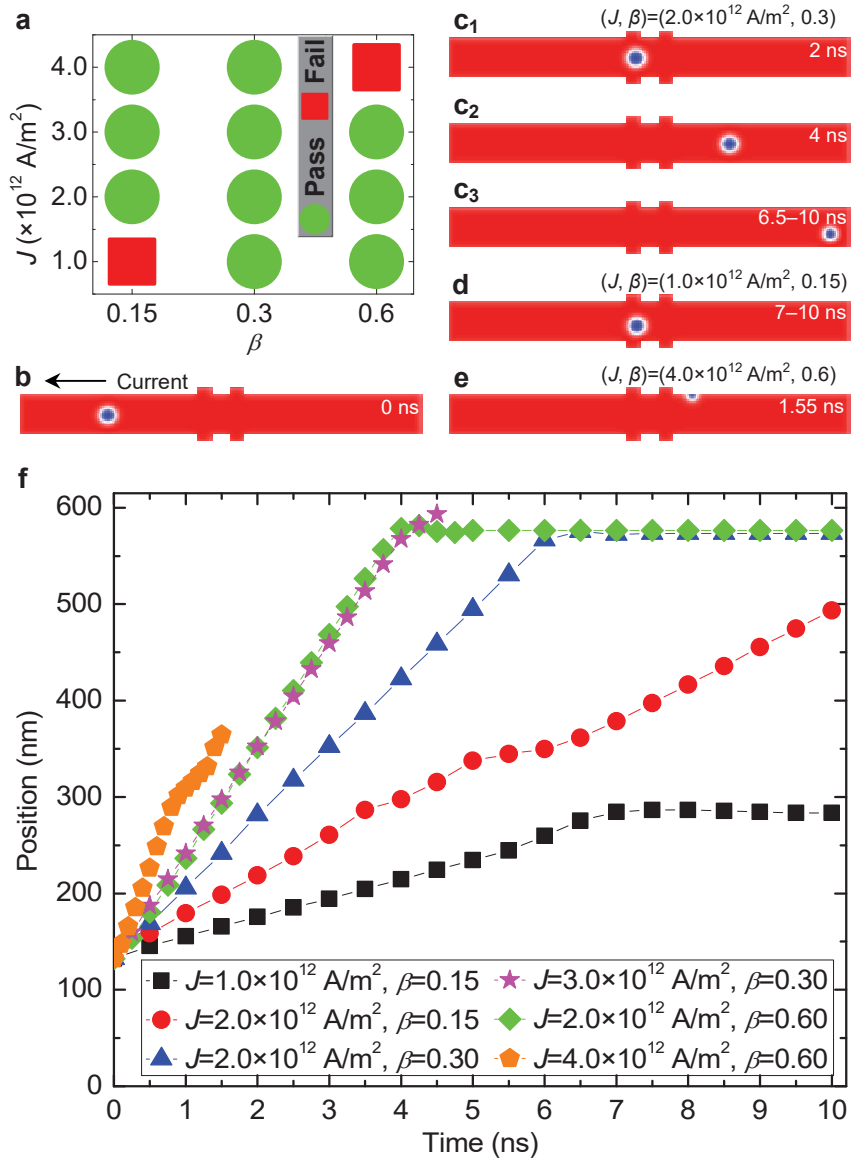

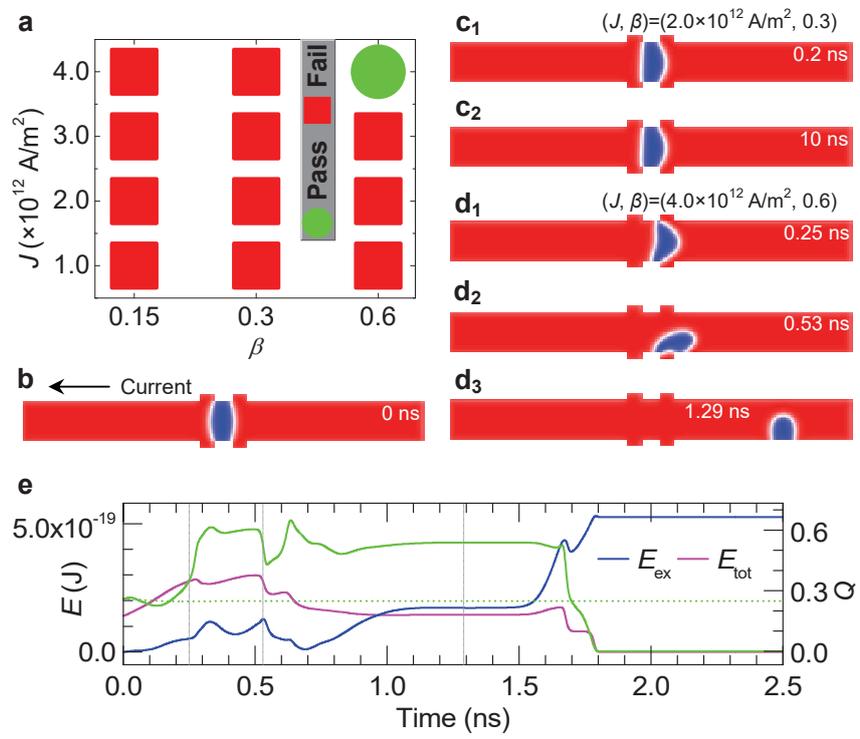

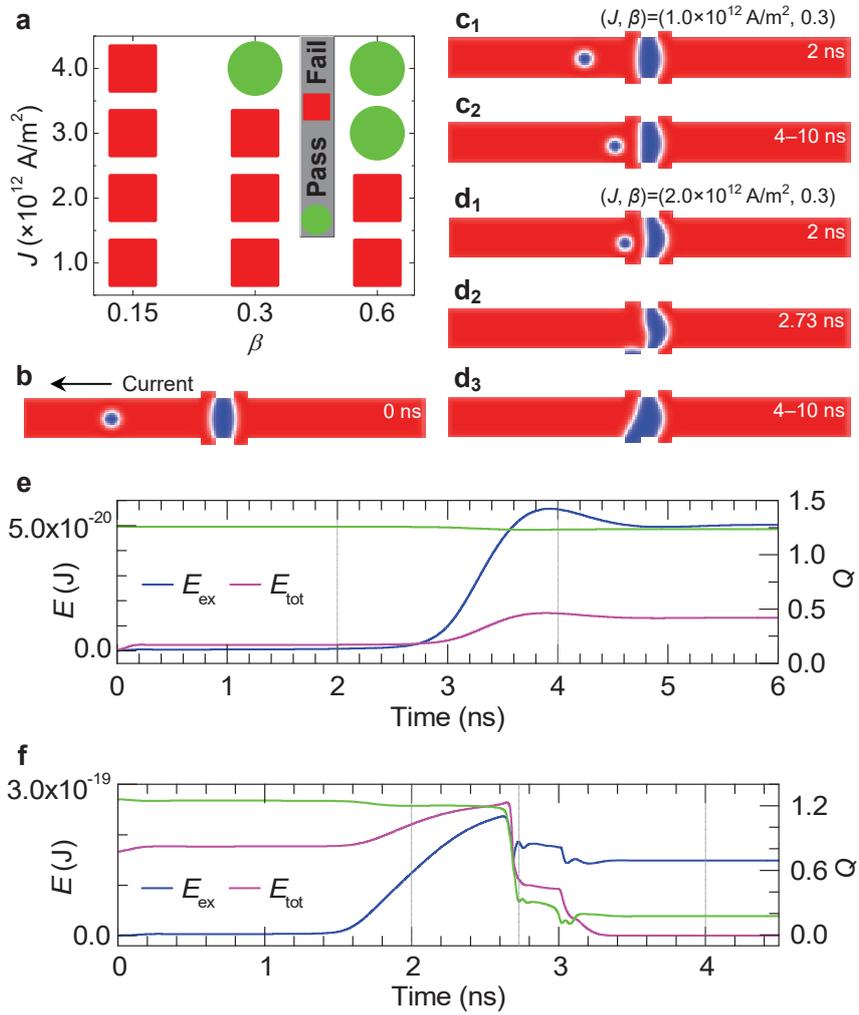

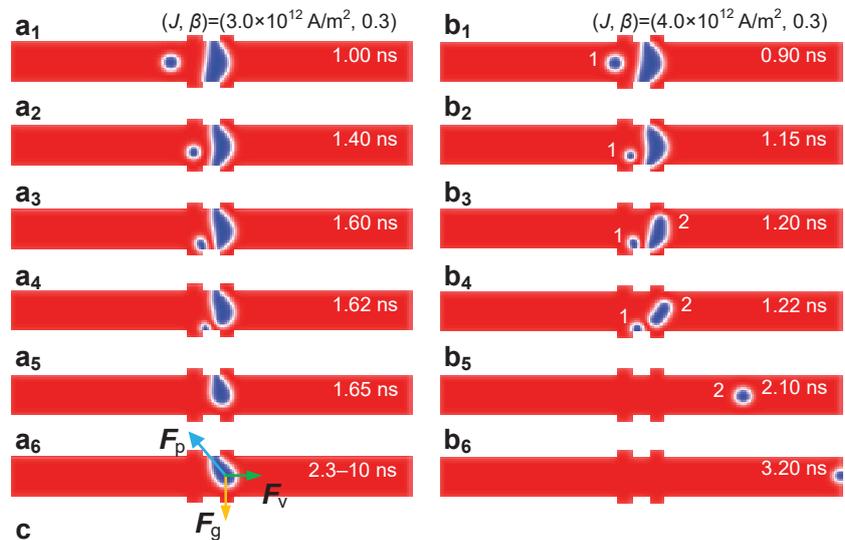
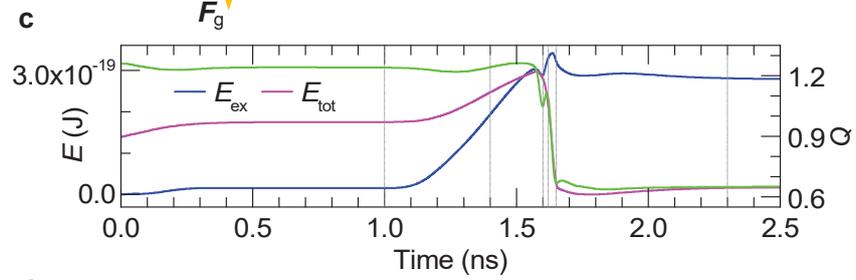
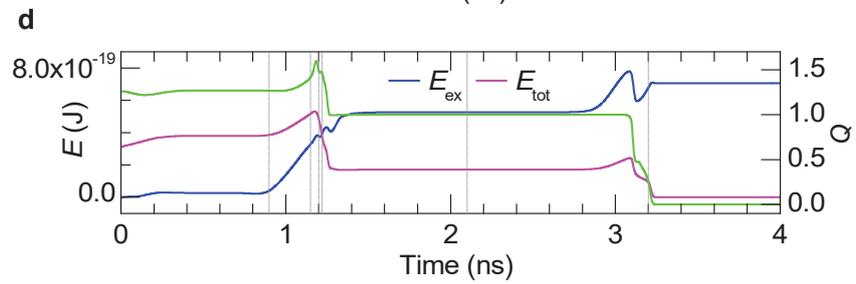

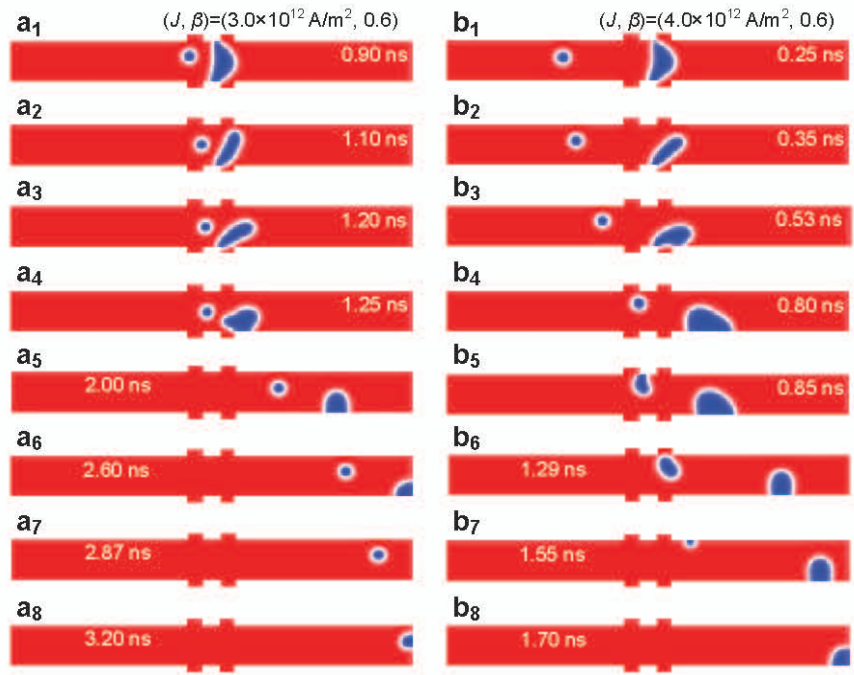
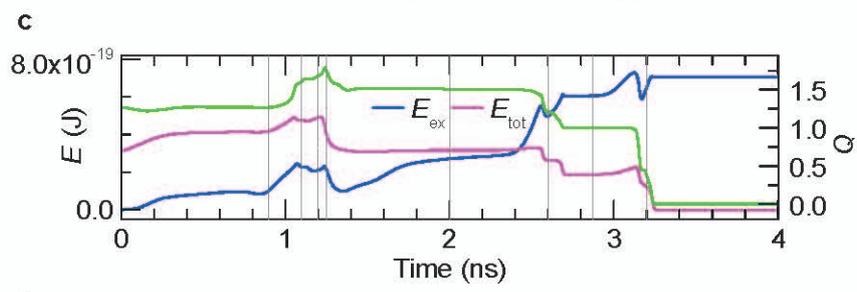
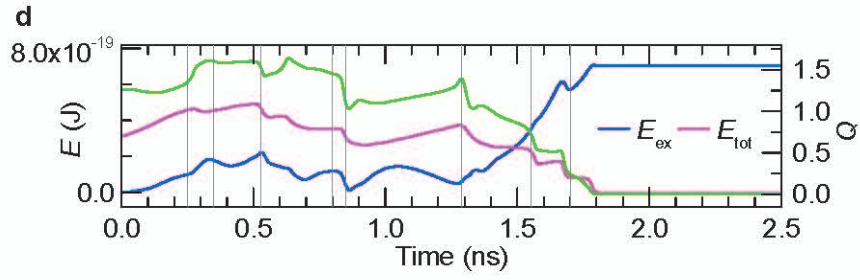

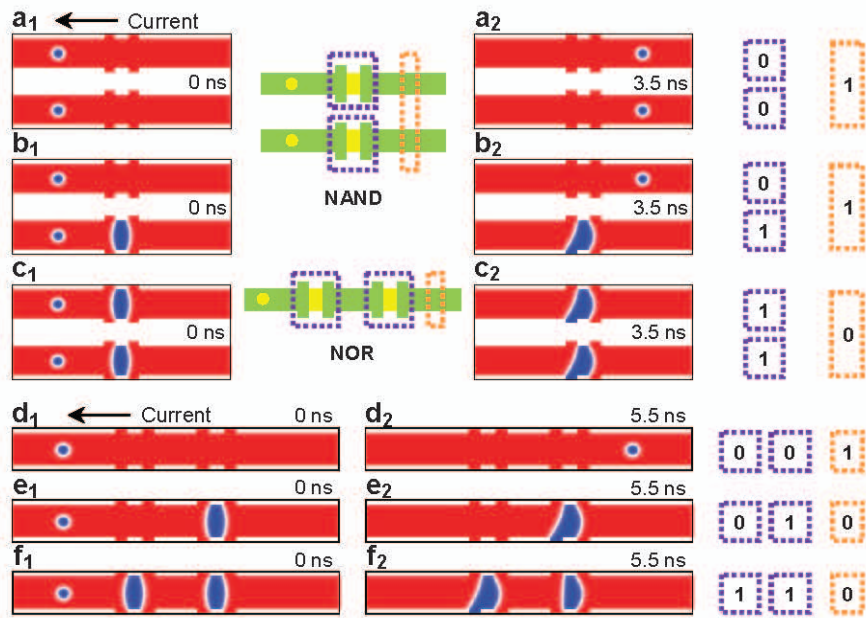